\documentclass[10pt,aps,pra,twocolumn,sort&compress,superscriptaddress,runinaddress,floatfix,preprintnumbers,showpacs,showkeys,nofootinbib]{revtex4-1}

\usepackage{dcolumn}
\usepackage{amssymb}
\usepackage{amsmath}
\usepackage{amsthm}
\usepackage{latexsym}
\usepackage{graphicx}
\usepackage{color}
\usepackage{ulem}
\usepackage{braket}
\usepackage{enumerate}

\newcommand{\hil}{\mathcal{H}}

\newcommand{\adag}{a^{\dagger}}
\newcommand{\abadag}{\bar{a}^{\dagger}}

\newcommand{\diste}{distinguishable }
\newcommand{\indiste}{indistinguishable }

\newcommand{\ketsub}[2]{\smash{\ket{#1}}^{}_{\! #2}}
\newcommand{\brasub}[2]{{}^{}_{#1}\!\smash{\bra{#2}}}
\newcommand{\braketsub}[3]{{}^{}_{#1}\!\smash{\braket{#2}}^{}_{\! #3}}

\newcommand{\tr}{\mathop{\textrm{tr}}}
\newcommand{\Mod}{\mathop{\textrm{mod}}}

\begin{document}

\preprint{ADP-12-31/T798}

\title{Uncertainty Relations and Indistinguishable Particles}
 

\author{Cael L.~Hasse}
\email{Electronic address: cael.hasse@adelaide.edu.au}
\affiliation{Special Research Centre for the Subatomic Structure of Matter 
and Department of Physics, University of Adelaide 5005, Australia.}
\date{\today}
\begin{abstract}
We show that for fermion states, measurements of any two finite outcome particle quantum numbers (e.g.\ spin) are not constrained by a minimum total uncertainty.  We begin by defining uncertainties in terms of the outputs of a measurement apparatus.  This allows us to compare uncertainties between multi-particle states of distinguishable and indistinguishable particles. Entropic uncertainty relations are derived for both distinguishable and indistinguishable particles. We then derive upper bounds on the minimum total uncertainty for bosons and fermions.  These upper bounds apply to any pair of particle quantum numbers and depend only on the number of particles $N$ and the number of outcomes $n$ for the quantum numbers.  For general $N$, these upper bounds necessitate a minimum total uncertainty much lower than that for distinguishable particles.  The fermion upper bound on the minimum total uncertainty for $N$ an integer multiple of $n$, is zero.  Our results show that uncertainty limits derived for single particle observables are valid only for particles that can be effectively distinguished. Outside this range of validity, the apparent fundamental uncertainty limits can be overcome.
\end{abstract}

\maketitle

\section{Introduction}

Quantum theory imposes fundamental limitations on our ability to simultaneously predict the outputs of measurements of different observables\footnote{A closely related subject is that of fundamental limitations quantum theory imposes on our ability to perform simultaneous non-demolition measurements of different observables.}. This observation was formalised by the Uncertainty Principle \cite{Heis27,Rob29}, which has played a pivotal role in the conceptual clarification of quantum theory. It also plays an important role in quantum information theory and cryptography \cite{Dam05,Wehn10,Bart10,Oppen10,Giov04}.

Most of the analysis of
uncertainty relations is done within the framework of distinguishable particles, important exceptions being the analysis of Bohr and Rosenfeld \cite{Bohr33} and the work of Glauber \cite{Glauber63}.  These exceptions explore uncertainty relations for field observables such as components of the electric and magnetic field and mode quadratures. The central goal of this paper is to derive and understand uncertainty relations for measurements of particle quantum numbers (q-numbers) on multi-particle states. Naively one may expect uncertainty relations for single particle states to extrapolate in a simple way. Is this true? Does a conventional understanding of single particle uncertainty relations ever break down for multi-particle states?

\section{Entropic Uncertainty}

Consider two q-numbers $\mathcal{A}$ and $\mathcal{B}$ for a particle, each with (finite) $n$ mutually exclusive possible outcomes. Consider the corresponding observables which we represent by operators $A_1$ and $B_1$ in an $n$-dimensional Hilbert space $\hil$. Let $\left\{\ket{i}\right\}$ and $\left\{\ket{\bar{j}}\right\}$ be the complete bases of eigenstates of $A_1$ and $B_1$ respectively. A good measure of the uncertainty of an observable is the Shannon entropy of its probability distribution. For a pure state $\ket{\psi} \in \hil$, the Shannon entropies for $\mathcal{A}$ and $\mathcal{B}$ are given by
\begin{align}
H(A_1;\psi) &:= -\sum^n_{i=1} P_i(A_1;\psi) \ln P_i(A_1;\psi), \\
H(B_1;\psi) &:= -\sum^n_{j=1} P_j(B_1;\psi) \ln P_j(B_1;\psi)
\end{align}
where $P_i(A_1;\psi) := \left|\braket{i | \psi}\right|^2$,
$P_j(B_1;\psi) := \left|\braket{\bar{j} | \psi}\right|^2$, and the
units are in nats (which we use throughout the paper).  Note
the maximum value of a Shannon entropy is given when one has a
constant probability distribution, i.e., $H(A_1;\psi)$ and
$H(B_1;\psi) \leq \ln n$.

The sum of these Shannon entropies is constrained by the uncertainty relation 
\cite{Kraus87,Maas88}:
\begin{equation}
H(A_1;\psi) + H(B_1;\psi) \geq -2 \ln c,
\label{smiley}
\end{equation}
where
\begin{equation}
c := \max_{j,k}\left|U_{jk}\right| \quad \textrm{and} \quad U_{jk} 
:= \braket{j | \bar{k}}. \label{eqn:c}
\end{equation}
By definition, $c$ is bounded: $0 \leq c \leq 1$. As Shannon
entropies for discrete probability distributions are positive
definite, for observables with $c < 1$, relation~(\ref{smiley})
implies at least one of the entropies must be non-zero.

We have characterized relation (\ref{smiley}) for operators $A_1$ and
$B_1$, which refer to particle q-numbers. However, the result applies
to any $s$-outcome operators acting on an $s$-dimensional Hilbert
space $\hil$ where $s \in \mathbb{N}$. We can therefore apply (\ref{smiley}) 
to multi-particle states. The distinguishability of particle ordering becomes 
important when one considers more than 1 particle. Does an extension of
(\ref{smiley}) to multi-particle states depend on whether one
considers distinguishable or indistinguishable particles?

\section{$N$-Particle Uncertainty Relation For \diste particles}

Consider $N$ \diste particles. There are $n$ possible outcomes
for \textit{each} particle, giving us $n^N$ possible outcomes in
total. Let $A^{(i)}$ be the operator representing the q-number $\mathcal{A}$ 
for the $i$\textsuperscript{th} particle with a basis of eigenstates
$\{\smash{\ket{m}}_i\}$ such that
\begin{equation}
A^{(i)}\ket{m}_{\!i} := \alpha_m \ket{m}_{\!i},
\end{equation}
where $\ket{m}_i \in \hil_i$ and $\hil_i$ is the Hilbert space for the 
$i$\textsuperscript{th} particle. The $N$-particle state $\ket{\psi_N}$ 
is then an element of $\bigotimes^N_{i=1}\hil_i$. Define a new observable 
whose eigenstates correspond to the distinct outcomes of a measurement of 
$\mathcal{A}$ on the system, where the measurement can distinguish
particle ordering:
\begin{equation}
A_N := \sum^n_{m_1, \ldots, m_N=1} e_{m_1\cdots m_N}
       \Bigg\{\bigotimes^N_{i=1} \ketsub{m_i}{i} \Bigg\}  
        \Bigg\{\bigotimes^N_{j=1} \brasub{j}{m_j} \Bigg\},
\end{equation}
where the eigenvalues $e_{m_1 \cdots m_N}$ are chosen to be non-degenerate. Then for an $N$-particle state $\ket{\psi_N}$, the Shannon entropy of $\mathcal{A}$ is given by
\begin{align}
H  ( &A_N;\psi_N) :=&  \nonumber \\
&-\sum_{i_1,\ldots,i_N=1}^n &\!\!\!\!\!P_{i_1\ldots i_N}(A_N;\psi_N)\ln P_{i_1\ldots i_N}(A_N;\psi_N),
\end{align}
where $P_{i_1\ldots
  i_N}(A_N;\psi_N)=\vert\brasub{1}{i_1}\cdots\brasub{N}{i_N}{\psi_N}\rangle\vert^2$.

Note that $H(A_N; \psi_N)$ is the joint Shannon entropy for the outcomes of $A^{(i)}$ for each particle. Define an analogous observable and entropy for $\mathcal{B}$. Utilizing relation~(\ref{smiley}) but replacing $A_1$ with $A_N$ and $B_1$ with $B_N$, one finds an entropic uncertainty relation
\begin{align}
H(A_N&; \psi_N) + H(B_N; \psi_N) \nonumber \\
&\geq -2\ln\Biggl\{\max_{\underset{r_1, \ldots, r_N}{\scriptscriptstyle m_1, \ldots, m_N}}
\Biggl|\bigotimes^N_{i=1} \brasub{i}{m_i} \bigotimes^N_{j=1} \ketsub{\bar{r}_{j}}{j} 
\Biggr|\Biggr\} \nonumber \\
&= -2\ln\Biggl\{\max_{\underset{r_1, \ldots, r_N}{\scriptscriptstyle m_1, \ldots, m_N}}
\Biggl|\prod^N_{i=1} \braketsub{i}{m_i | \bar{r}_i}{i} \Biggr|\Biggr\} \nonumber \\
&= -2\ln c^N \nonumber \\
&=-2N\ln c
\label{grin}
\end{align}
where $c$ is defined as in Eq.~(\ref{eqn:c}).

This result is easily understood for cases where $\ket{\psi_N}$ is
unentangled:  
\begin{equation}
\ket{\psi^u_N} = \ketsub{\psi^{(1)}}{i} \cdots \ketsub{\psi^{(N)}}{N} \,.
\end{equation}
The average information gained about $A_N$ over the ensemble equals 
the sum of the average information gained about $A^{(i)}$ for each
particle:
\begin{align}
H(A_N; \psi^u_N) &= \sum_{i=1}^N H(A^{(i)}; \psi^{(i)}) \,, \nonumber \\
H(B_N; \psi^u_N) &= \sum_{i=1}^N H(B^{(i)}; \psi^{(i)})\,. \label{:<}
\end{align}
Then,
\begin{align}
H(A_N&; \psi^u_N) + H(B_N; \psi^u_N) \nonumber \\ &= \sum_{i=1}^N \left( H(A^{(i)}; \psi^{(i)}) + H(B^{(i)}; \psi^{(i)}) \right) \nonumber \\
&\ge -2N \ln c, \label{:)|}
\end{align}
which gives the same bound on $H(A_N;\psi_N) + H(B_N;\psi_N)$ as relation~(\ref{grin}) for the special case where $\ket{\psi_N}$ is unentangled.

Relation~(\ref{:)|}) highlights an interesting feature of relation~(\ref{grin}); that it is unaffected by possible entanglement of $\ket{\psi_N}$. It is not immediately obvious to us why this is the case.  Our thoughts go as follows:
\begin{enumerate}[(a)]
\item The state of particle `$i$', $\rho^{(i)} = \tr_{\neq i} [ \ket{\psi_N} \bra{\psi_N}]$, can become mixed, i.e. the von Neumann entropy $S(\rho^{(i)}) = -\tr [ \rho^{(i)} \ln \rho^{(i)} ] > 0$. A non-zero $S(\rho^{(i)})$ constrains the Shannon entropy of $\mathcal{A}$ and $\mathcal{B}$ for particle `$i$' \cite{Hasse12,Schumacher96}:
\begin{equation}
\left.\begin{array}{c} H(A^{(i)}, \rho^{(i)}) \\[0.5mm] 
                       H(B^{(i)}, \rho^{(i)}) \end{array}\right\}  \ge S(\rho^{(i)}) \,.
\end{equation}
This suggests it may be difficult for entangled states to saturate the bound (\ref{grin}).
\item The subadditivity of the Shannon entropy implies
\begin{equation}
H(A_N, \psi_N) < \sum_{i=1}^N H(A^{(i)}, \rho^{(i)}),
\end{equation}
for correlated observables $A^{(i)}$, and similarly for $\mathcal{B}$.  This disallows a direct use of our previous observation in explaining why entangled states cannot overcome relation (\ref{:)|}).
\end{enumerate}

The salient feature of both relations~(\ref{grin}) and (\ref{:)|}) is 
the proportionality to $N$ of the lower bounds. We claim that this feature 
is indicative of the assumption that it is possible to define probabilities 
for the outcomes of measurements of q-numbers $\mathcal{A}$ and $\mathcal{B}$ 
for \textit{each} particle.

It is well known \cite{Maas88} that, for a given quantum state, relation~(\ref{smiley}) 
is not necessarily the strongest possible bound. Accordingly, neither 
are relations~(\ref{grin}) and (\ref{:)|}).

\section{Comparing distinguishable and indistinguishable particles}

To compare uncertainties of particle q-numbers for \diste and \indiste
particles, we have to be careful about the precise meaning of the
probabilities of the various outcomes for multi-particle states.

For \diste particles, the meaning is simple. For a 2 particle state we
can consider observables for each particle $\mathcal{O}^{(1)}$ and
$\mathcal{O}^{(2)}$ such that
\begin{equation}
[\mathcal{O}^{(1)},\mathcal{O}^{(2)}]=0.
\end{equation}
This defines the natural tensor product decomposition
$\hil_1\otimes\hil_2$ where $\hil_i$ corresponds to the $i$-th
particle. Suppose we have a state $\rho$ which is pure and unentangled
\begin{equation}
\rho = \ketsub{\psi^{(1)}}{1}\ketsub{\psi^{(2)}}{2}\brasub{2}{\psi^{(2)}}\brasub{1}{\psi^{(1)}},
\end{equation}
where $\ketsub{\psi^{(1)}}{1}\in\hil_1$ and
$\ketsub{\psi^{(2)}}{2}\in\hil_2$. Let $\mathcal{O}^{(1)}$ and
$\mathcal{O}^{(2)}$ be position projectors for positions $x_1$ and $x_2$ respectively, then
\begin{equation}
\langle\mathcal{O}^{(1)}\mathcal{O}^{(2)}\rangle = |\psi^{(1)}(x_1)|^2|\psi^{(2)}(x_2)|^2,
\end{equation}
which is the probability for particle 1 to be measured at $x_1$ and particle 2
to be measured at $x_2$.

Consider \indiste particles. Now all observables commute with
permutations of the particle states {\it i.e.}, for a permutation
operator $P_{1\leftrightarrow 2}$ that permutes the states in $\hil_1$
and $\hil_2$ and any observable $\mathcal{O}$,
\begin{equation}
[P_{1\leftrightarrow 2},\mathcal{O}]=0.
\label{permcomm}
\end{equation}
This defines a superselection rule to constrain direct measurements
to irreducible representations of the permutation group {\it i.e.},
symmetric and anti-symmetric under $P_{1\leftrightarrow 2}$.

Importantly for our considerations, the constraint (\ref{permcomm})
means {\it all} observables must act non-locally on the tensor
product decomposition\footnote{So the Fock space is a much better
  decomposition for the states.} $\hil_1\otimes\hil_2$. The observable
which then corresponds closest to
$\mathcal{O}^{(1)}\mathcal{O}^{(2)}=\ketsub{x_1}{1}\brasub{1}{x_1}\otimes\ketsub{x_2}{2}\brasub{2}{x_2}$
is then,
\begin{equation}
\mathcal{O}^{(12)}:=\frac{1}{2}\big(\ketsub{x_1}{1}\ketsub{x_2}{2}\pm\ketsub{x_2}{1}\ketsub{x_1}{2}\big)\big(\brasub{1}{x_1}\brasub{2}{x_2}\pm\brasub{1}{x_2}\brasub{2}{x_1}\big),
\end{equation}
with $+$ or $-$ corresponding to bosons or fermions respectively. The expectation value 
of this observable gives us the probability that we shall measure a particle at 
$x_1$ and another at $x_2$.

The probabilities given by $\langle\mathcal{O}^{(1)}\mathcal{O}^{(2)}\rangle$
and $\langle\mathcal{O}^{(12)}\rangle$ have fundamentally different
meanings. Predictions for \diste particles are {\it counterfactual} \cite{Perez93}
because they assume the existence of particle labels that cannot be
measured.

The probabilities and hence Shannon entropies in
relation~(\ref{smiley}) are given physical meaning by the standard
axioms of quantum theory that the outcomes of a measurement
(eigenstates of an observable) correspond to outputs of a measuring
apparatus with probabilities for occurrence defined in the normal
way. What is of physical interest are these outputs. We assume: 
\begin{enumerate}[(a)]
\item An ideal apparatus whose outputs are in one-to-one
  correspondence with the outcomes.
\item Meaningful probabilities can be defined for these outcomes. 
\end{enumerate}
Then a {\it Shannon entropy can be defined on the probability
  distribution of the outputs}. Assumption (a) ensures direct
usefulness to physics. This Shannon entropy of the outputs rather than
the outcomes (eigenstates of some observable) can then be used to
compare different theories whose probabilities have very different meanings.

Suppose we perform measurements of q-numbers $\mathcal{A}$ and $\mathcal{B}$ on an
infinite, homogeneous ensemble of pure $N$-particle states. Define the
Shannon entropy of the outputs of the measurement apparatus as
$H_N(\mathcal{A})$ and $H_N(\mathcal{B})$ for measurements of $\mathcal{A}$ and $\mathcal{B}$,
respectively. 

Identifying the outputs with outcomes for measurements of $N$ \diste
particles gives us, 
\begin{equation}
  H_N(\mathcal{A}) = H(A_N;\psi_N),
\end{equation}
and similarly for $\mathcal{B}$.

For the outputs of measurements of $\mathcal{A}$ and $\mathcal{B}$ over 
$N$-particle states we can write
\begin{equation}
H_N(\mathcal{A}) + H_N(\mathcal{B}) \geq \chi_{\textrm{min}}(N;\mathcal{A},\mathcal{B}),
\end{equation}
where $\chi_{\textrm{min}}(N; \mathcal{A}, \mathcal{B})$ is the real minimum 
over the $N$-particle Hilbert space. In this context, the uncertainty relation 
for distinguishable particles gives
\begin{equation}
\chi_{\textrm{min}}^{\textrm{dist}}(N;\mathcal{A},\mathcal{B}) = -2N\ln c.
\label{eqn:circle}
\end{equation}

\section{$N$-Particle Uncertainty Relation for \indiste particles}

We now define $H_N(\mathcal{A})$ and $H_N(\mathcal{B})$ for $N$-particle states of
\indiste particles.

In order to define $H_N(\mathcal{A})$, the type of measurement interaction must
be considered.  Consider particles with two compatible q-numbers $\mathcal{A}$
and $\mathcal{C}$, with eigenmodes $\alpha_i$ and $c_j$ respectively.  Let $m$
be the number of $\mathcal{C}$ eigenmodes and consider the case where $m >
n$. The two-particle states $\ket{\alpha_1, c_1; \alpha_2, c_2}$ and
$\ket{\alpha_2, c_1; \alpha_1, c_2}$ are reliably distinguishable in an
experiment measuring $\mathcal{A}$ if the interaction also couples to modes of
$\mathcal{C}$. The modes of $\mathcal{C}$ then act as a reference frame \cite{Bart07,
  Eisert00, Bartlett03, Jones05, Jones06} for the effective ordering distinguishability of the
particles. 

As an example, $\mathcal{C}$ may relate to the position of the
particle. Measurements can then distinguish particle ordering if the
measurement interaction has position dependence, such as the magnetic
field in a Stern-Gerlach experiment being localized in space.

For an $N$-particle state, \textit{the $n^N$ possible outputs we have for \diste particles correspond\footnote{This correspondence can be given by the isomorphism $f_\chi$ defined in \cite{Sasaki:2011}.} to \indiste particle states where the modes of $\mathcal{C}$ are different for each particle and also fixed over the set of outputs}.  These states---where the modes of $\mathcal{C}$ are different for each particle---form only a subspace of the $N$-particle states for \indiste particles. It is not guaranteed that the other $N$-particle states satisfy the same bounds for $H_N(\mathcal{A}) +H_N(\mathcal{B})$.

We require the states of \diste particles to correspond to states of
\indiste particles with modes of $\mathcal{C}$ fixed for the following reasons.
With \diste particles, for states that are elements of
$\otimes_{i=1}^{N}\mathcal{H}_i$ and measurements of $\mathcal{A}$ or $\mathcal{B}$, it
doesn't matter what the q-number $\mathcal{A}$ is referenced to.  With \indiste particles where
the reference mode $\mathcal{C}$ is modelled, different reference modes are
distinguished; i.e., the states $|\alpha_1, c_1; \alpha_2, c_2\rangle$
and $|\alpha_1,c_3; \alpha_2, c_4\rangle$ are different but give
equivalent probabilities for $\mathcal{A}$ and $\mathcal{B}$.  Thus they can be identified
with the same state $|1\rangle_1 |2\rangle_2$.

Consider particle creation operators $\adag_{jl}$ (fermionic or
bosonic) which create a particle in mode `$j$' for $\mathcal{A}$ and mode `$l$'
for $\mathcal{C}$ such that the state $\adag_{jl}\ket{0}$ (where $\ket{0}$ is
the Fock vacuum) corresponds\footnote{They are equivalent in terms of
  giving the same probabilities for particle q-numbers.  The Fock
  space however, allows for measurements of q-numbers which are
  impossible for a quantum theory with \diste particles, i.e.,
  superpositions of occupation number eigenstates.} to the $A_1$ eigenstate
$\ket{j,l}^{}_{1}$. One can also define ladder operators
$\abadag_{il}$ such that $\abadag_{il}\ket{0}$ corresponds to an
eigenstate of $B_1$, $\ket{\bar{i},l}^{}_{1}$. The two sets of
operators are related by the Bogoliubov transformation
\begin{equation}
\abadag_{il} := \sum_{k=1}^{n} U_{ik} \adag_{kl}. \label{bog}
\end{equation}

For $N$-particle states of \indiste particles, the Shannon entropy of the
outcomes of $\mathcal{A}$ is constructed in the following way. Define the
observable
\begin{align}
A_{N|C} &:= \sum_{i_{11},\ldots,i_{nm}=0}^{N} f_{i_{11}\cdots i_{nm}} \delta_{(\sum_{b,c}i_{bc})N}\nonumber \\
&\times \left\{\prod^{n,m}_{j=1,l=1} \frac{\big(\adag_{jl}\big)^{i_{jl}}}{\sqrt{i_{jl}!}}\ket{0}\right\}
\left\{\bra{0}\prod^{n,m}_{k=1,r=1} \frac{\big(a_{kr}\big)^{i_{kr}}}{\sqrt{i_{kr}!}}\right\}\,,
\end{align}
where $ \delta_{(\sum_{b,c}i_{bc})N}$ is a Kronecker delta. The
eigenvalues $f_{i_1\cdots i_n}$ are, chosen to be non-degenerate and
the ordering of the ladder operators arbitrary. (The ordering of
ladder operators does not affect any result of this paper.  In cases
where the ordering isn't specified, the reader is free to choose it.)
Then the corresponding Shannon entropy is
\begin{align}
&H  ( A_{N|C};\phi) =  \nonumber \\
&-\sum_{i_{11},\ldots,i_{nm}=0}^n P_{i_{11}\ldots i_{nm}}(A_{N|C};\phi)\ln P_{i_{11}\ldots i_{nm}}(A_{N|C};\phi),
\end{align}
where
\begin{equation}
P_{i_{11}\ldots i_{nm}}(A_{N|C};\phi) =
\delta_{(\sum_{b,c}i_{bc})N}\Bigl|
  \bra{0}\prod^{n,m}_{k=1,l=1}\frac{(a_{kl})^{i_{kl}}}{\sqrt{i_{kl}!}}\ket{\phi}\Bigr|^2.
\end{equation}
Identifying outputs to outcomes gives us
\begin{equation}
H_N(\mathcal{A}) = H(A_{N|C}; \phi)
\end{equation}
for a state of $N$ \indiste particles $\ket{\phi}$ and an apparatus that couples to $\mathcal{A}$ and $\mathcal{C}$. An analogous entropy for B can also be defined with an observable $B_{N|C}$ which is given by $A_{N|C}$ with the ladder operators $\adag_{jl}$ replaced by $\abadag_{jl}$.

We are now in a position to construct an uncertainty relation for $H_N(\mathcal{A})$ and $H_N(\mathcal{B})$. We choose the Hilbert space to be the $N$-particle subspace of the entire Fock space of these modes. Applying relation~(\ref{smiley}) to $A_{N|C}$ and $B_{N|C}$ instead of $A_1$ and $B_1$, we obtain
\begin{align}
H(&A_{N|C};\phi) + H(B_{N|C};\phi) \geq \nonumber \\ 
&-2\ln\Biggl\{\,\max_{\underset{j_{11}, \ldots, j_{nm}}{\scriptscriptstyle i_{11}, \ldots, i_{nm}}}\,
\biggl|\delta_{(\sum_{b,c}i_{bc})N} G(i, j) \biggr|\Biggr\},
\label{shock}
\end{align}
where
\begin{equation}
G(i, j) := \braket{0 | \prod^{n,m}_{k=1,l=1} \frac{(a_{kl})^{i_{kl}}}{\sqrt{i_{kl}!}
}\prod^{n,m}_{t=1,r=1} \frac{(\abadag_{tr})^{j_{tr}}}{\sqrt{j_{tr}!}} | 0}.
\end{equation}

So how does (\ref{shock}) compare to (\ref{eqn:circle})? Firstly, a
sanity check. If we restrict ourselves to an $N$-particle subspace
where the modes of $\mathcal{C}$ are different for each particle and also fixed,
the outcomes and probabilities of a measurement of $\mathcal{A}$ or $\mathcal{B}$ should
be in one-to-one correspondence with the \diste particles case. Thus, with this
restriction, $H(A_{N|C}; \psi) + H(B_{N|C}; \psi)$ should never be
smaller than $-2N \ln c$. We can compute the RHS of (\ref{shock}) with
this restriction. The overlap between eigenstates of $A_{N|C}$ and
$B_{N|C}$ can be given without loss of generality by
\begin{align}
G(i, j) &= \braket{0 | a_{i_1 1}\cdots a_{i_N N} \bar{a}^\dagger_{j_1 1} \cdots \bar{a}^\dagger_{j_N N} | 0} \nonumber \\
&= U_{i_1 j_1} \cdots U_{i_N j_N}.
\end{align}
Thus with this restricted subspace,
\begin{equation}
H(A_{N|C}; \psi) + H(B_{N|C}; \psi) \ge -2N \ln c,
\end{equation}
which agrees with relation~(\ref{grin}).

For the full $N$-particle Hilbert space, relation (\ref{shock}) can be written 
explicitly in terms of $U_{ij}$.  Denote $P_N$ as the group of permutations of 
the components of the elements of $\mathbb{N}^N$.  For example, let 
$\tilde{k} = (k_1,\cdots,k_N) \in \mathbb{N}^N$ and $\sigma \in P_N$ such that 
$\sigma$ swaps the first and second components:
\begin{equation}
\sigma\tilde{k} = (k_2,k_1,k_3,\cdots,k_N)\,.
\end{equation}

Define
\begin{align}
V(\tilde{k},\tilde{t}) &= U_{t_1k_1}\cdots U_{t_Nk_N}\,, \\
W(\tilde{k},\tilde{k}') &= \delta_{k_1'k_1} \cdots\delta_{k_N'k_N} \,,
\end{align}
and 
\begin{align}
M(\tilde{k},\tilde{t},\tilde{l}) &= \Bigg\{ \sum_{\sigma'\in P_N} W(\sigma'\tilde{k},\tilde{k}'=\tilde{k}) W(\sigma'\tilde{l},\tilde{l}'=\tilde{l}) \Bigg\}^{\frac{1}{2}} \notag \\
&\times \Bigg\{\sum_{\sigma''\in P_N} W(\sigma''\tilde{t},\tilde{t}'=\tilde{t}) W(\sigma''\tilde{l},\tilde{l}'=\tilde{l}) \Bigg\}^{\frac{1}{2}}\,,
\end{align}
where $k_i, t_i \leq n$ and $l_i \leq m$.  Also define
\begin{align}
G^{\mathrm{boson}}(\tilde{k},\tilde{t},\tilde{l}) &= \frac{1}{M}\sum_{\sigma\in P_N} V(\sigma\tilde{k},\tilde{t}) W(\sigma\tilde{l},\tilde{l}'=\tilde{l}
)\,, \notag \\
G^{\mathrm{ferm}}(\tilde{k},\tilde{t},\tilde{l}) &= \sum_{\sigma \in P_N} \mathrm{sgn}(\sigma) V(\sigma\tilde{k},\tilde{t}) W(\sigma\tilde{l},\tilde{l}'=\tilde{l})\,.
\end{align}
Then for bosons and fermions respectively, relation (\ref{shock}) gives us
\begin{equation}
\chi^{\mathrm{boson}}_{\mathrm{min}}(N; \mathcal{A},\mathcal{B}) = -2 \ln \left| 
\max_{\tilde{k},\tilde{l}, \tilde{t}}G^{\mathrm{boson}} \right| \label{eqn:flatface}
\end{equation}
and
\begin{equation}
\chi^{\mathrm{ferm}}_{\mathrm{min}}(N; \mathcal{A},\mathcal{B}) = -2 \ln \left| 
\max_{\tilde{k},\tilde{l}, \tilde{t}}G^{\mathrm{ferm}} \right|\,. \label{eqn:frownface}
\end{equation}

We can immediately compare these uncertainty relations to Eq.~(\ref{eqn:circle}).  
Eq.~(\ref{eqn:circle}) has two nice properties.  Firstly, 
it is proportional to $N$.  Secondly, different choices of $\cal A$ and 
$\cal B$ affect only the slope of this proportionality.  In contrast, the 
euations above for bosons and fermions cannot be expected to have these 
properties.  This is because they involve sums of $V(\tilde{k},\tilde{t})$ 
which generally are complex numbers and thus generally interfere with each other.

Can we say anything general about the N dependence of $\chi^{\mathrm{boson}}_{\mathrm{min}}$ 
and $\chi^{\mathrm{ferm}}_{\mathrm{min}}$?  Surprisingly we can.  In the next two sections we 
derive upper bounds on them, dependent only on $N$ and $n$.

\section{Bosons}

Let us focus on states that do not correspond to states of \diste
particles. Any interesting differences between (\ref{eqn:flatface}) and
(\ref{eqn:circle}) are likely to show up in this region of the Hilbert
space. Consider for instance an $N$-boson eigenstate of $A_{N|C}$
where every particle is in the same $\mathcal{C}$ eigenmode `$l$':
\begin{equation}
|\omega\rangle := |\hat{\omega}\rangle/\sqrt{\langle\hat{\omega} | \hat{\omega}\rangle} 
\mbox{, where }
\ket{\hat{\omega}} := a^\dagger_{i_n l} \cdots a^\dagger_{i_N l} \ket{0}.
\end{equation}
By construction, $H(A_{N|C}; \omega) = 0$. An upper bound can be found for 
$H(B_{N|C}; \omega)$ using the following argument:
\begin{equation}
H(B_{N|C}; \omega) \le \ln ( \textrm{no. possible outcomes} ).
\end{equation}

The Bogoliubov transformation (\ref{bog}) does not change (a) the number of 
particles, and (b) the $\mathcal{C}$ eigenmodes. Each component of $\ket{\omega}$ 
in the $\mathcal{B}$-mode basis must have $N$-particles where each particle is 
still in $\mathcal{C}$ eigenmode `$l$'. How many possible components with these 
properties are there? The question is isomorphic to the question, ``how many ways 
can I place $N$ indistinguishable balls into $n$ distinguishable boxes?'' The 
answer \cite{Constantine87} is $\binom{N+n-1}{N}$. Thus,
\begin{equation}
H(A_{N|C}; \omega) + H(B_{N|C}; \omega) \le \ln \binom{N+n-1}{N}. \label{:[]}
\end{equation}
For $N \ge n$,
\begin{equation}
\ln \binom{N+n-1}{N} = (n-1) \ln N - \sum_{k=1}^{n-1} \ln k + O(1/N),
\end{equation}
and hence this upper bound is of order $\ln N$, which shows that 
Eq.~(\ref{eqn:flatface}), unlike Eq.~(\ref{eqn:circle}), cannot be proportional 
to $N$.

\begin{figure}[tb]
\includegraphics{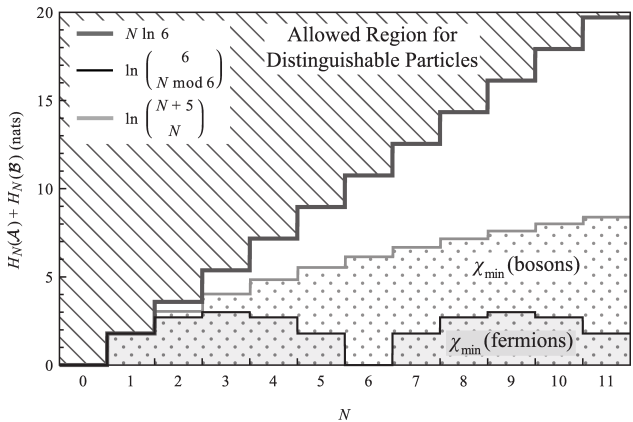}
\caption{\label{fig:plot}Comparison of bounds on uncertainties $H_N(\mathcal{A})$ 
and $H_N(\mathcal{B})$ for \diste particles (\ref{eqn:circle}) and \indiste 
particles (relation (\ref{eqn:triangle}) for fermions and relation 
(\ref{eqn:square}) for bosons) for varying particle number $N$. The q-numbers 
are chosen to be complementary with $n=6$ distinct outcomes. This gives 
$c = 1/\sqrt{6}$.}
\end{figure}

Relation (\ref{:[]}) provides an upper bound on $\chi_\textrm{min}$
\begin{equation}
\chi_\textrm{min}^\textrm{boson}(N;\mathcal{A},\mathcal{B}) 
  \leq \ln \binom{N + n -1}{N}\,,
\label{eqn:square}
\end{equation}
which is dependent only on $N$ and $n$.  This is demonstrated in Fig.\ \ref{fig:plot}.

To illustrate relation~(\ref{eqn:square}), consider an $N=2=n$ example with photons, given by the $A_{2|C}$ eigenstate
\begin{equation}
\ket{\gamma\gamma} := a^\dagger_{1k} a^\dagger_{2k} \ket{0},
\end{equation}
where 1 and 2 denote polarization eigenmodes of $\mathcal{A}$ and $k$ denotes an eigenmode of $\mathcal{C}$.  The q-number $\mathcal{C}$ could be for instance position, momentum, or a combination of angular momentum q-numbers. Choose eigenmodes of $\mathcal{B}$ as rotations of 1 and 2 through an angle $\alpha$,
\begin{align}
&\quad a^\dagger_{1k} = \cos \alpha a^\dagger_{\alpha_1 k} + \sin \alpha a^\dagger_{\alpha_2 k}, \nonumber \\
\textrm{and} \nonumber \\
&\quad a^\dagger_{2k} = -\sin \alpha a^\dagger_{\alpha_1 k} + \cos \alpha a^\dagger_{\alpha_2 k},
\end{align}
such that
\begin{align}
\ket{\gamma \gamma} &= \sqrt{2}\cos\alpha \sin\alpha (\tfrac{1}{\sqrt{2}}a^\dagger_{\alpha_2 k} a^\dagger_{\alpha_2 k} - \tfrac{1}{\sqrt{2}}a^\dagger_{\alpha_1 k} a^\dagger_{\alpha_1 k})\ket{0} \nonumber \\
&+ (\cos^2\alpha - \sin^2\alpha)a^\dagger_{\alpha_1 k} a^\dagger_{\alpha_2 k}|0\rangle.
\end{align}
Suppose $\mathcal{B}$ is complementary to $\mathcal{A}$, i.e. $\alpha = \pi/4$; then,
\begin{equation}
H(A_{2|C}; \gamma\gamma) + H(B_{2|C}; \gamma\gamma) = \ln 2,
\end{equation}
which is smaller than the RHS of (\ref{eqn:square}) and {\it half} the lower bound
(\ref{eqn:circle}) for $H_N(\mathcal{A}) + H_N(\mathcal{B})$ with \diste particles. Interestingly, the upper
bound (\ref{eqn:square}) can be saturated for $\alpha = \frac{1}{2} \arctan \sqrt{2}$
where $\mathcal{B}$ is {\it not} complementary to $\mathcal{A}$:
\begin{equation}
H(B_{2|C} |_{\alpha = \frac{1}{2} \arctan \sqrt{2}}; \gamma\gamma) = \ln 3.
\end{equation}

\section{Fermions}

Let's turn our attention to fermions, whose ladder operators we shall denote 
by $b_{jl}$ and $b^\dagger_{jl}$. For the case $N=n$, the RHS of 
Eq.~(\ref{eqn:frownface}) can be computed in the following way. Consider 
the state
\begin{equation}
\ket{\lambda_f} := b^\dagger_{1l}\cdots b^\dagger_{nl}\ket{0}
\end{equation}
which is an eigenstate of $A_{n|C}$. Rewriting the state in terms of modes 
of $\mathcal{B}$ gives
\begin{align}
\ket{\lambda_f} &= \sum^n_{i_1,\ldots,i_n=1}U^{-1}_{1i_1}\cdots U^{-1}_{ni_n} \bar{b}^\dagger_{i_1l} \cdots \bar{b}^\dagger_{i_nl}\ket{0} \nonumber \\
&= \sum^n_{i_1,\ldots,i_n=1}U^{-1}_{1i_1}\cdots U^{-1}_{ni_n} \epsilon_{i_1\cdots i_n} \bar{b}^\dagger_{1l} \cdots \bar{b}^\dagger_{nl}\ket{0} \nonumber \\
&= \det \left(U^{-1}\right) \bar{b}^\dagger_{1l} \cdots \bar{b}^\dagger_{nl}\ket{0},
\label{eqn:wide1}
\end{align}
where we also used the fact that $\ket{0}$ is invariant under Bogoliubov 
transformations (\ref{bog}) that do not mix creation and annihilation operators. 
Thus
\begin{equation}
\left|\braket{0 | \prod^n_{k=1} b_{kl} \prod^n_{r=1} \bar{b}^\dagger_{rl} | 0}\right| 
= \left|\det \left(U^{-1}\right)\right| = 1
\end{equation}
which is the maximum possible value for the overlap between the bases of 
eigenstates of $A_{n|C}$ and $B_{n|C}$. Hence $\chi^\textrm{ferm}_\textrm{min}$
is {\it zero} for $N = n$. This is in stark contrast to the 
\diste $n$-particle uncertainty relation that constrains 
$H_n(\mathcal{A}) + H_n(\mathcal{B}) \ge -2 n \ln c$.

By considering the exclusion principle, we can find more states that circumvent 
Eq.~(\ref{eqn:circle}): Fermions are constrained to be in orthogonal modes. 
These are modes of the product of all compatible q-numbers of the particles. 
The only way to create two or more particles in mode `$i$' for q-number 
$\mathcal{A}$ is for the $\mathcal{C}$ modes of the particles to be orthogonal. 
Consider again an $N$ ($\leq n$)-particle state where every particle is in the 
same $\mathcal{C}$ mode,
\begin{equation}
\ket{\lambda'_f; i_1, \ldots, i_N} := \prod^N_{j=1} b^\dagger_{i_j k} \ket{0}. \label{eqn:lamprime}
\end{equation}
The q-number $\mathcal{C}$ does not act as a reference frame for $\mathcal{A}$. 
There are only $\binom{n}{N} \leq n^N$ allowed choices for the labels $i_j$ 
which correspond to different outcomes for a measurement of $\mathcal{A}$. 
This also applies to $\mathcal{B}$. Following the same arguments we made for 
bosons, a limit on the number of outcomes puts a limit on the Shannon entropy 
of the probability distribution over the outcomes; for a generic normalized 
superposition of the states $\ket{\lambda'_f; i_1, \ldots, i_N}$ with no 
reference frame for $\mathcal{A}$
\begin{equation}
\ket{\lambda'_f} = \sum_{i_1, \ldots,i_N} c_{i_1\cdots i_N} \ket{\lambda'_f; i_1,\ldots, i_N},
\end{equation}
the entropies for the outcomes of $\mathcal{A}$ and $\mathcal{B}$ are constrained 
from above, by the exclusion principle:
\begin{align}
H(A_{N|C};\lambda'_f) &\leq \ln \binom{n}{N}, \label{eqn:haqft} \\
H(B_{N|C};\lambda'_f) &\leq \ln \binom{n}{N}. \label{eqn:hbqft}
\end{align}
The extreme case of these constraints is when $N = n$ such that both entropies 
must be zero. This is equivalent to the results obtained for the state 
$\ket{\lambda_f}$.

The constraints (\ref{eqn:haqft}) and (\ref{eqn:hbqft}) can be extended to arbitrary $N = an + b$ where $b < n$. This is done by combining the states $\ket{\lambda_f}$ and $\ket{\lambda'_f}$:
\begin{equation}
\ket{\lambda''_f} = \sum_{i_i, \ldots, i_b} c_{i_1 \cdots i_b} \prod_{j = 1}^b b_{i_j k_a}^\dagger \prod_{l = 1}^{a - 1} \prod_{r = 1}^n b^\dagger_{rk_l} \ket{0}, \label{eqn:lambdapp1}
\end{equation}
where $k_1, \ldots, k_a$ label orthogonal modes of $\mathcal{C}$. Then,
\begin{align}
H(A_{N|C}; \lambda''_f) &\le \ln \binom{n}{N \Mod n}, \label{eqn:newa}\\
H(B_{N|C}; \lambda''_f) &\le \ln \binom{n}{N \Mod n}. \label{eqn:newb}
\end{align}

If one has $c_{i_1\cdots i_b} = 1$ for some choice of $i_1,\ldots,i_b$ and the other choices all zero, then $H(A_{N|C}; \lambda''_f) = 0$ without changing (\ref{eqn:newb}). Thus we arrive at an upper bound on $\chi_\textrm{min}$,
\begin{equation}
\chi_\textrm{min}^\textrm{ferm}(N;\mathcal{A},\mathcal{B}) \leq \ln \binom{n}{N \Mod n}\,.
\label{eqn:triangle}
\end{equation}

As an example of the extreme case $n=N$, where $H_N(\mathcal{A}) = 0 
= H_N(\mathcal{B})$, consider a dineutron state where both neutrons are in 
an $s$-wave and their spins are anti-aligned on the $z$-axis. The q-number 
$\mathcal{A}$ will correspond to these spins on the $z$-axis. Denote 
$\adag_{\uparrow q}$ and $\adag_{\downarrow q}$ as the creation operators 
for the up and down $s$-wave neutrons with all other q-numbers given by $q$. 
The state is given by 
\begin{equation}
\ket{nn} := b^\dagger_{\uparrow q} b^\dagger_{\downarrow q}\ket{0}. \label{eqn:nn}
\end{equation}

For a single spin half particle in a pure state, there exists only {\it one} 
axis where the particle has definite spin. For $\ket{nn}$, both particles 
have definite spin in {\it every} direction; consider $\mathcal{B}$ modes 
as modes of spin rotated from the $z$-axis by $\vec{\theta}$. Let 
$U_{ij} = (\exp\{i\vec{\theta}\cdot\vec{\sigma}\})_{ij}$, where 
$\vec{\sigma} = (\sigma_1,\sigma_2,\sigma_3)$ is the Pauli matrix vector 
such that 
\begin{equation}
\bar{b}^\dagger_{\theta_i q} = \sum_{j=\uparrow,\downarrow}
   \left(\exp\{i\vec{\theta}\cdot\vec{\sigma}\}\right)_{ij} b^\dagger_{jq}\,.
\end{equation}
Then
\begin{equation}
\ket{nn} = \bar{b}^\dagger_{\theta_{\uparrow q}} \bar{b}^\dagger_{\theta_{\downarrow q}}\ket{0} \quad \forall \hspace{2mm} \vec{\theta}\,.
\end{equation}

Another two particle example is a deuteron state. Consider both nucleons to be in an $s$-wave and their spins aligned. The q-numbers $\mathcal{A}$ and $\mathcal{B}$ with $H_N(\mathcal{A}) = 0 = H_N(\mathcal{B})$ correspond to choices of isospin basis. Measurements of superpositions of a proton and neutron would have to overcome the charge superselection rule \cite{Aharonov67, Dowling06}.

\section{Remarks}

In this paper, we derive three $N$-particle uncertainty relations: 
Eq.~(\ref{eqn:circle}) for \diste particles and (\ref{eqn:flatface})
and (\ref{eqn:frownface}) for indistinguishable bosons and fermions 
respectively. We develop an understanding of how particle q-number 
ordering information manifests itself with indistinguishable particles. 
Not all states of \indiste particles have effective distinguishability 
for the q-number of interest. This leads us to upper bounds 
(\ref{eqn:square}) and (\ref{eqn:triangle}) for $\chi_\textrm{min}$ 
(the minimum of $H_N(\mathcal{A})+H_N(\mathcal{B})$), which shows that 
$\chi_\textrm{min}$ for bosons or fermions is generally much smaller than 
$\chi_\textrm{min}$ for distinguishable particles. These upper bounds are 
due to the structure of indistinguishable particles and are only dependent 
on the number of particles $N$ and the number of outcomes $n$ for 
$\mathcal{A}$ and $\mathcal{B}$. That is, they are not dependent on 
the compatibility of $\mathcal{A}$ and $\mathcal{B}$.

Relation~(\ref{eqn:triangle}) in particular has an infinite number of zeroes, 
which occur when $N$ is an integer multiple of $n$. This means that fermions 
are not bound by a minimum total uncertainty for finite $n$.

Different experimental situations have to be considered separately. One 
situation is when the measurement interaction does not couple to any 
reference modes. The minimum total uncertainty $\chi_\textrm{min}$ will
generally be smaller than the lower bound given by Eq.~(\ref{eqn:circle}).
This can be considered akin to coarse graining where we are not interested
in all the information the state has to offer, such as particle ordering 
information. For the other situation, where the measurement interaction
does couple to reference modes, any loss of effective particle distinguishability
occurs at a fundamental level: ordering information cannot be accessed
irrespective of the capabilities of the measurement apparatus.

Our analysis shows that in the wider context of multi-particle states, the 
Heisenberg uncertainty limits {\it must} be reunderstood.  Apparent fundamental 
limitations on the precision of measurements of particle q-numbers can in 
principle be overcome.

We emphasize several points:
\begin{enumerate}[(a)]
\item  The no coupling case has a maximum $\binom{N+n-1}{N}$ outputs, which
  is the number obtained by simply ignoring particle ordering information. 
  We have not explored bounds on $H_N(\mathcal{A})+H_N(\mathcal{B})$ for
  \diste particles with particle ordering ignored. For this situation, there 
  is an upper bound on the minimum of $H_N(\mathcal{A}) + H_N(\mathcal{B})$ 
  which is the same as what we derived for bosons.
  Simply ignoring particle ordering will still give different results
  compared to bosons. The differences between these
  two situations are explored in \cite{Hasse:later}.

\item The definition of $A_{N|C}$ is symmetric under a swap between
  $\mathcal{A}$ and $\mathcal{C}$, i.e. $A_{N|C} = C_{N|A}$. Each state we considered in
  this paper had definite particle number for each mode of $\mathcal{C}$. If one
  considers states where this is not the case, $H(A_{N|C}; \phi) =
  H(C_{N|A}; \phi)$ is a quantification of the uncertainty in both $\mathcal{A}$
  and $\mathcal{C}$.

\item For ease of exposition, we described states such as
  $\ket{\omega}$ and $\ket{\lambda''_f}$ in terms of the labels of the
  ladder operators. For such states, where one loses a reference frame
  for $\mathcal{A}$ and $\mathcal{B}$, the identity of the particles for measurements of
  $\mathcal{A}$ and $\mathcal{B}$ is obscured and independence of the outcomes for these
  q-numbers becomes ambiguous. For instance, consider again the state
  $\ket{\lambda_f}$ where $H(A_{N|C}; \lambda_f) = 0 = H(B_{N|C};
  \lambda_f)$. One is tempted to say that \textit{every particle has
    definite values for $\mathcal{A}$ and $\mathcal{B}$ simultaneously}.

  The particles in the various modes of $\mathcal{A}$ and $\mathcal{B}$ are
  `elements of physical reality' according to the EPR criterion
  \cite{Einstein:1935rr,Ghirardi:2002}. However it is meaningless to say whether a particle in a
  certain mode of $\mathcal{A}$ is simultaneously in a mode of $\mathcal{B}$. The
  particles have lost every label which might distinguish one from the
  other. We consider the circumvention of relation~(\ref{grin}) by
  states such as $\ket{\omega}$ and $\ket{\lambda''_f}$ as the
  manifestation of this lack of distinguishability.

\item In many situations, the very concept of a particle becomes difficult 
  to define. The context of our arguments are less restrictive than in 
  relativistic quantum field theory where the Hilbert space is defined by 
  asymptotically non-interacting multiparticle states. We have assumed the 
  Fock space of a non-interacting field theory can define all multiparticle 
  states. We have also implicitly assumed that there exists particle q-number 
  measurements that can be performed on all interacting multiparticle states.

  Particles must have a long enough lifetime
  relative to the interaction time such that the states can be
  considered approximate energy eigenstates.  The dineutron for
  example is not stable \cite{PDG}.

  One must also consider that approximate energy eigenstates where particle
  fields begin to overlap will often not be given as simple products
  of ladder operators. For instance, multiple electrons bound in an
  atom interact with each other leading to energy eigenstates that are
  not simply products of ladder operators that create single electron
  energy eigenstates.  
  
  These are some of the factors that must be considered if one attempts to experimentally reach the new limits found in this paper.
\end{enumerate}

\begin{acknowledgments}
I thank R.~J.~Crewther and Lewis C.~Tunstall for useful comments and suggestions.  I also acknowledge technical support from Benjamin J.~Menadue, S.~Underwood and Dale S.~Roberts.  This work is supported by the Australian Research Council.
\end{acknowledgments}

\end{document}